\documentclass[manuscript]{acmart}

\usepackage{balance,bm}
\usepackage{xcolor,colortbl}
\usepackage{url}
\usepackage{enumitem}

\usepackage{graphicx}  
\usepackage{textcomp, soul}
\usepackage{wrapfig}
\usepackage{subcaption}
\usepackage{mathtools}
\usepackage{algorithmic}
\usepackage[ruled]{algorithm2e}
\usepackage[T1]{fontenc}
\usepackage[utf8]{inputenc}
\usepackage{multirow, verbatim, xspace}
\usepackage[many]{tcolorbox}
\usepackage{titlesec}
\usepackage{arydshln}
\usepackage{longtable}
\usepackage{booktabs}

\definecolor{linkColor}{RGB}{6,125,233}
\definecolor{green}{rgb}{0.0, 0.65, 0.31}
\definecolor{bleudefrance}{rgb}{0.19, 0.55, 0.91}
\definecolor{ceruleanblue}{rgb}{0.16, 0.32, 0.75}
\definecolor{grey}{HTML}{969696}
\definecolor{lightgrey}{HTML}{d7d7d7}
\definecolor{greybackground}{HTML}{e9ecef}
\definecolor{violet}{HTML}{6a51a3}
\definecolor{lgreen}{HTML}{5ab4ac}
\definecolor{dgreen}{HTML}{005a32}
\definecolor{purple}{HTML}{ae017e}
\definecolor{orange}{HTML}{d95f0e}

\definecolor{generate}{HTML}{9e3dc5}
\definecolor{understand}{HTML}{3dc40f}
\definecolor{evaluate}{HTML}{ff733c}

\definecolor{individual1}{HTML}{F5E4E4}
\definecolor{individual2}{HTML}{C95E5E}
\definecolor{care1}{HTML}{CEE9D9}
\definecolor{care2}{HTML}{417C59}
\definecolor{info1}{HTML}{EFDFF3}
\definecolor{info2}{HTML}{9249A3}
\definecolor{tech1}{HTML}{D9E2FC}
\definecolor{tech2}{HTML}{435993}

\colorlet{tableheadcolor}{gray!25} 
\colorlet{tablerowcolor}{gray!15} 
\colorlet{tablerowcolor2}{gray!12} 
\colorlet{tablerowcolor3}{gray!25} 

\renewcommand{\textrightarrow}{$\rightarrow$}

\colorlet{tableheadcolor}{gray!25} 
\colorlet{tablerowcolor}{gray!5} 

\newtcolorbox{roundBox}{
  enhanced,
  breakable,
  before skip = 0.5em,
  after skip = 0.5em, 
  top = 0.3em,
  bottom = 0.3em,
  colback = grey!10, 
  boxrule = 0pt,
  rounded corners,
  arc = 3pt,
  fontupper= \small\itshape
}

\newtcolorbox{featureBox}{
  enhanced,
  breakable,
  before skip = 0.5em,
  after skip = 0.5em, 
  top = 0.3em,
  bottom = 0.3em,
  colback = greybackground, 
  boxrule = 0pt,
  rounded corners,
  arc = 3pt,
}

\AtBeginDocument{%
  \providecommand\BibTeX{{%
    \normalfont B\kern-0.5em{\scshape i\kern-0.25em b}\kern-0.8em\TeX}}}

\newif{\ifhidecomments}
\hidecommentsfalse 
\ifhidecomments
   \newcommand{\jiawei}[1]{}
   \newcommand{\amy}[1]{}
   \newcommand{\darshi}[1]{}
   \newcommand{\laura}[1]{}
   \newcommand{\munmun}[1]{}  
\else
   \newcommand{\jiawei}[1]{\textbf{\sffamily{\textcolor{lgreen}{[#1 -- Jiawei]}}}}
   \newcommand{\amy}[1]{\textbf{\sffamily{\textcolor{olive}{[#1 -- Amy]}}}}
   \newcommand{\darshi}[1]{\textbf{\sffamily{\textcolor{orange}{[#1 -- Darshi]}}}}
   \newcommand{\laura}[1]{\textbf{\sffamily{\textcolor{mediumblue}{[#1 -- Laura]}}}}
   \newcommand{\munmun}[1]{\textbf{\sffamily{\textcolor{purple}{[#1 -- Munmun]}}}}

\acmConference[CSCW'25]{CSCW'25 Beyond Information Workshop}{October 18--22, 2025}{Bergen, Norway}
\begin{document}

\title{Large Language Models as Information Sources: Distinctive Characteristics and Types of Low-Quality Information}

\author{Jiawei Zhou}
\affiliation{
  \institution{Georgia Institute of Technology}
  \city{Atlanta}
  \state{GA}
  \country{USA}
}
\email{j.zhou@gatech.edu}

\author{Amy Z. Chen}
\affiliation{
  \institution{Georgia Institute of Technology}
  \city{Atlanta}
  \state{GA}
  \country{USA}
}
\email{amychen@gatech.edu}

\author{Darshi Shah}
\affiliation{
  \institution{Georgia Institute of Technology}
  \city{Atlanta}
  \state{GA}
  \country{USA}
}
\email{dshah435@gatech.edu}

\author{Laura M. Schwab-Reese}
\affiliation{
  \institution{Purdue University}
  \city{West Lafayette}
  \state{IN}
  \country{USA}
}
\email{lschwabr@purdue.edu}

\author{Munmun De Choudhury}
\affiliation{
  \institution{Georgia Institute of Technology}
  \city{Atlanta}
  \state{GA}
  \country{USA}
}
\email{munmund@gatech.edu}

\renewcommand{\shortauthors}{Jiawei Zhou et al.}

\begin{abstract}  

Recent advances in large language models (LLMs) have brought public and scholarly attention to their potential in generating low-quality information. While widely acknowledged as a risk, low-quality information remains a vaguely defined concept, and little is known about how it manifests in LLM outputs or how these outputs differ from those of traditional information sources. In this study, we focus on two key questions: What types of low-quality information are produced by LLMs, and what makes them distinct than human-generated counterparts? We conducted focus groups with public health professionals and individuals with lived experience in three critical health contexts (vaccines, opioid use disorder, and intimate partner violence) where high-quality information is essential and misinformation, bias, and insensitivity are prevalent concerns. We identified a typology of LLM-generated low-quality information and a set of distinctive LLM characteristics compared to traditional information sources. Our findings show that low-quality information extends beyond factual inaccuracies into types such as misprioritization and exaggeration, and that LLM affordances fundamentally differs from previous technologies. This work offers typologies on LLM distinctive characteristics and low-quality information types as a starting point for future efforts to understand LLM-generated low-quality information and mitigate related informational harms. We call for conceptual and methodological discussions of information quality to move beyond truthfulness, in order to address the affordances of emerging technologies and the evolving dynamics of information behaviors.

\end{abstract}



\maketitle

\section{Introduction}

Low-quality information in the forms of false information, hate speech, and other misleading narratives has become a long-standing societal concern with far-reaching consequences on individual mental well-being~\cite{saha2019prevalence, Verma2022ExaminingHealth}, exacerbated racial discrimination~\cite{He2021RacismCrisis, Islam2020COVID-19RelatedAnalysis} and even hate crimes~\cite{kopytowska2017stereotypes, Yam_2022}, and threats to social cohesion~\cite{harel2020normalization}. The consequences can be severe during health crisis, a time when people are experiencing heightened anxiety and fear and emotionally venerable to the flourishing of problematic information~\cite{Setbon2010FactorsA/H1N1, Verma2022ExaminingHealth, nelson2020danger}: preventable hospital admissions and mortality~\cite{Coleman2020HundredsMisinformation, Islam2020COVID-19RelatedAnalysis} and eventually eroded trust in health systems~\cite{rodriguez2021public}.

This proliferation of low-quality information is partially rooted in the nature of today's information ecosystems, where information and communication technologies (ICTs) allows anyone to not only consume but also produce and promote information with ease and speed. While these affordances have democratized information access and authorship~\cite{zhao2017consumer, costello2016impact}, they also provide problematic content with an expansive outlet to disseminate misleading or biased ideas~\cite{sylvia2020we}. Compounding this is the transformed way we interact with information. As worried by the post-truth era fear~\cite{mcintyre2018post}, online users do not always verify sources or assess information reliability~\cite{hansen2003adolescents, hassoun2023practicing}, while algorithmic curation that rewards engagement over accuracy~\cite{} and echo chambers can further amplify biased perspectives~\cite{jiang2021social}. In such contexts, the authority of traditional knowledge institutions, such as scientific bodies, public health agencies, and professional journalism, can be diminished, making it easier for problematic content to gain traction. And this prevalence of low-quality information can gradually erode trust and discourage reliable information seeking behaviors~\cite{zhao2017consumer, augustaitis2021online}.

Now large language models (LLMs) have entered this already fragile information ecosystems, as both powerful tools for information access and potential amplifiers of disorder. Unlike earlier information technologies such as traditional search engines, LLMs do not retrieve and present existing documents; they generate new text in a probabilistic manner based on vast and opaque training data~\cite{Brown2020LanguageLearners,vaswani2017attention}. This generative capacity enables highly fluent, seemingly authoritative responses to a limitless range of queries. However, it also means that LLM outputs may combine accurate and inaccurate information or frame facts in ways that subtly mislead~\cite{zhou2023synthetic, chang2024survey, liang2022holistic}. Their conversational format and stylistic polish make them more persuasive than many traditional sources, even when the underlying content is flawed~\cite{zhou2023synthetic}. The result is a broad class of risks that can be less visible than overt misinformation yet potentially more pervasive.

In this work, we focus on the well-documented but still vaguely defined potential risk of LLM-generated low-quality information. While this issue is increasingly acknowledged in public and scholarly discourse, it currently lacks clear conceptual boundaries and comprehensive empirical evidence, making it difficult to study and address systematically. This gap is consequential: as a concept broader than misinformation or hate speech, low-quality information does not necessarily hinge on falsity or malicious intent; thus, its harm may arise from distorting how readers interpret, prioritize, or emotionally respond to information. Moreover, the same attributes that make LLMs compelling, such as linguistic fluency and authoritative style, also make them uniquely capable of producing and scaling informational harm at a speed and scale we have never seen before. Thus, in this paper, we asks two questions: \textit{What types of low-quality information are produced by LLMs? What makes LLM-generated low-quality information distinct from, and in some cases riskier than, other human creations of its kind?}

We conducted focus group sessions with ten public health professionals and ten individuals with lived experience who have actively sought online information, in order to jointly explore the impacts using LLMs for health informational needs. 
We selected vaccines, opioid use disorder, and intimate partner violence as topics for different sessions based on their significance across different dimensions of public health: infectious disease prevention, chronic and well-being care, and community health and safety --- all demanding high-quality information with existing prevalent issues such as misinformation, biases, and sensitivity. The results are 1) a typology of LLM-generated low-quality information grounded in empirical data and domain expertise, and 2) a list of distinctive characteristics of LLMs compared to traditional information sources that contribute to the production and persistence of potential informational harms.

By situating LLM-generated low-quality information within the broader information disorder, we highlight the additional challenges brought by LLMs and argue that the community's conceptual and methodological discussions need to expand beyond factual accuracy.

\section{Context} 

To contextualize and evaluate the risks of using large language models (LLMs) for informational needs in public health, we chose three distinct and critical issues: vaccines, opioid use disorder (OUD), and intimate partner violence (IPV). The selection of these topics was driven by their significance across different dimensions of public health --- infectious disease prevention, chronic and well-being care, and community health and safety. Each topic underscores the importance of high-quality information in public health communication, which often involves both sensitive or high-stakes issues and diverse populations. 
Specifically, vaccines play a crucial role in infectious disease prevention, while misinformation and public distrust significantly contribute to vaccine hesitancy~\cite{pierri2022online, puri2020social, diamond2022polyvocality}. Effective counter-speech is needed to combat these misconceptions and enhance trust~\cite{zhang2021cultural, zimet2013beliefs}. OUD is a major issue in chronic and well-being care, which is exacerbated by stigma~\cite{elsherief2024identification, olsen2014confronting} and vulnerability~\cite{yamamoto2019association, van2020socioeconomic}. This health crisis needs public education to reduce biases and create a supportive environment. IPV is a high-risk and stigmatized issue in community health and safety that demands highly sensitive and supportive communication~\cite{schwab2017attitudinal, mittal2024news}. IPV involves complex challenges, including personal safety, mental health, and legal and financial challenges~\cite{schwab2016associations, freed2017digital}. Providing high-quality, empathetic information is essential for supporting survivors and addressing their comprehensive needs.

\section{Method} 

\subsection{Participants and Recruitment}

We conducted focus groups with two participant groups: professional experts (E1–E10) and members of the general public (P1–P10). Professional experts are people work in the public health sector, such as social workers, nurses, community health workers, and researchers, all of whom have experience working on at least one of the three topics. The general participants are individuals who have actively sought information about one of these three issues. Additionally, participants in the sessions of OUD and IPV are individuals who have lived experience with these issues. 

Professionals were recruited via snowball sampling through public health departments, professional organizations, alliances, shelters, hospital and university networks, and researchers specializing in relevant health communication topics, with outreach spanning multiple U.S. states. Public participants were recruited on the Prolific~\footnote{https://www.prolific.com/} research platform. 

Our screening surveys collected demographics, occupation, AI attitudes, familiarity with LLMs, and availability to suggested focus group timings. Participants were selected and invited for focus groups solely based on their availability. Ten professionals and ten public participants participated in focus groups. On average, our participants have a moderate level of previous experience in AI tools and usage of LLM tools, with health professionals having slightly more experience in AI but less use of LLM.

\subsection{Study Procedure}

All six focus groups were conducted virtually on Zoom to accommodate participants from across the U.S. To provide a safe and comfortable environment for our participants, we held separate sessions for professionals and experiencers to ensure that experiencers participated in sessions with individuals who shared similar lived experiences. 

Each session began with an overview of goals, participant rights, and introductions. We first discussed participants' current practices in seeking or sharing health information, challenges in assessing quality, precautions in creating and sharing information, and common misconceptions about the selected topics. Then, we provided a brief introduction to LLMs, emphasizing their probabilistic text generation and potential for inaccuracy, followed by demonstrations using example health-related questions adapted from real social media posts. Participants were encouraged to suggest their own questions to illustrate varying response styles. To illustrate variation in interaction, questions were tested separately on ChatGPT and Google AI Overview, with a clear emphasis that the purpose was not to compare the tools directly. Participants discussed the quality of the outputs using a think-aloud approach. Lastly, they brainstormed potential negative consequences of using LLMs for health information.

\subsection{Data Analysis}

We analyzed transcripts and brainstorming notes using reflexive thematic analysis~\cite{braun2006using, braun2019reflecting, braun2021one}. Three researchers independently coded a subset of sessions to identify instances of low-quality information and LLM-specific risk factors, iteratively refining codes and organizing them into higher-level categories. The first author then coded the remaining sessions, with ongoing team discussions to resolve ambiguities. Drawing on both our engagement with participants and our interpretation of the data, the final outputs were (1) a typology of LLM-generated low-quality information and (2) a set of distinctive LLM characteristics compared to traditional information sources.

\subsection{Privacy and Ethics}
This study was approved by the Institutional Review Board (IRB) at our institution. The demographic information and video recordings were collected with consent and later anonymized. We refrained from collecting any personally identifiable information from people with lived experience, and data from screened-out or dropped-out participants was discarded.
Throughout the recruitment and focus groups, we assured the participants that their participation was completely voluntary, all questions were optional, and their responses would be anonymous. 

\section{Findings}

\subsection{Types of Low-Quality Information Generated by LLMs}
\label{low_quality}

We present a typology of LLM-generated low-quality information, grounded in empirical data and participant input. These types capture diverse ways in which LLM outputs may fall short of supporting informational needs. 

\begin{itemize}[leftmargin=1.5em]
    \item \textbf{Misinformation:} Misinformation is false or partially false information that may be created with or without the intent to deceive and can be spread either intentionally or unintentionally.  
    
    \item \textbf{Dangerous advice:} Dangerous advice refers to guidance or recommendations that pose a risk of harm to individuals or groups, potentially leading to unsafe actions or negative well-being outcomes. This is more concerning in critical situations, such as emergency treatments and responses to health crises, where dangerous advice can lead to severe harm or even loss of life. For example, in IPV issues, LLM's advice of setting boundaries may sound reasonable on the surface level but in reality could invoke abusers’ more severe harm to the survivors. Our participants explained that, \textit{``setting clear boundaries could be really dangerous thing to do is to let your partner, this abusive partner, know `Hey. I'm gonna set boundaries now', and that can trigger a whole another form of reaction and abuse.''} (E7).

    \item \textbf{Oversimplified answers:} Health information rarely comes in a one-serves-all answer and demands a personal understanding of their prioritization, genetics, medical, and family history, as \textit{``one thing that works for one person might not work for another''} (P4).

    \item \textbf{Omission of critical details:} Critical omissions refer to the failure to include essential information that is necessary for a full understanding of a situation, decision, or recommendation. In public health, omitting key details can be just as harmful as spreading misinformation, as it can lead to incomplete understanding, poor decision-making, and inadequate responses to risks. For instance, IPV experts noted that in handling situations involving violence, the need for a safety plan is always a priority, and omitting this information is considered harmful.

    \item \textbf{Exaggeration:} Exaggeration happens when information is amplified beyond its actual significance or truthfulness. This can include overstating the severity, benefits, or risks associated with a health issue or treatment. This distortion can misinform public perceptions and lead to harmful behaviors. Overemphasizing benefits, as in prescribing opioids, may cause risks to be overlooked, while overstating risks, such as police inaction in violence cases, can foster fear and avoidance of help-seeking.

    \item \textbf{Biased statements:} Biased statements tend to present information in a way that favors one perspective, group, or outcome over others. Biases have long been an issue in healthcare that comes in forms such as medical racism, victim blaming, and stigma of certain diseases. This issue becomes more concerning when people mistakenly perceive technology as neutral and bias-free, while in reality, it can perpetuate or even amplify existing biases. 

    \item \textbf{Outdated or non-representative conclusions:} These are statements that are either based on outdated data or fail to accurately reflect the diversity and complexity of the population. Therefore, LLM generations may overlook recent advancements in knowledge or disregard the variability in individual or community experiences. As explained by our participant, \textit{``they (tech companies) are going to use data that's not reflective of what's happening right now because they don't have it. So a real negative consequence is that, assuming that the datasets are up to date, timely, accurate […] And I think, in public health that is not the case at all.''} (E5). 

    \item \textbf{Contradictory or confusing statements:} Contradictory or confusing statements tend to present conflicting or unclear information and cause misunderstandings, which cam can cause hesitation in seeking care or mismanagement of health conditions, and may result in harmful outcomes. In interacting with LLMs, their outputs may vary or contain contradictions or hallucinations across different interactions. 

    \item \textbf{Incorrect prioritization of information:} Incorrect prioritization of information occurs when content is arranged or emphasized in a way that does not accurately reflect its relative importance or relevance to an individual’s needs or a specific context. This can result in a loss of trust or hesitation in seeking help, and may lead to ineffective or insufficient support. Participants noted that the list format often used in LLM responses may imply a ranked importance of information, even though this is not intentional or by design. Even if prioritization were intentional, accurately determining the order of needs often requires years of experience working with the community.

\end{itemize}

\subsection{Distinctive LLM Characteristics Compared to Traditional Information Sources}



We summarize a list of characteristics that contribute to the identified risks and differentiate LLMs from traditional health information sources. 
We acknowledge that some characteristics may not apply to all models and product deployments --- such as for fine-tuned or specialized models and emerging interaction modes --- and mark these characteristics with asterisks (\text{*}).

\begin{itemize}[leftmargin=1.5em]
    \item \textbf{Probability-based without understanding the content:} LLMs are probability-based models that predict the conditional probability of the next token without truly understanding language. Thus, output can vary between generations with no guarantee of information quality or sensitivity to emotions and cultures.
    
    \item \textbf{Untraceable information sources:} Unlike traditional information sources where authors, affiliations, and citations are provided to readers to provide credibility indicators, users are unable to trace the origins of LLM-generated information (without other techniques).
    
    \item \textbf{Positioned to serve general purposes\text{*}:} Most LLMs claim to serve general purposes, but they tend to lack domain-specific knowledge and practice-based expertise.

    \item \textbf{Generation quality dependent on training data quality:} LLM outputs are only as good as the training data, which raises questions about whether the training data is representative, fair, up-to-date, and accurate.
    
    \item \textbf{Standardized and formal linguistic style\text{*}:} LLMs have a standardized and formal language style that can sound authoritative or unempathetic. Their outputs are typically presented in a list format, which implies comprehensiveness and ranked importance in the provided answers.

    \item \textbf{Users role in initiating and forming prompts\text{*}:} When interacting with LLMs, users need to be proactive in initiating conversations and forming appropriate questions to convey crucial details and contexts.

    \item \textbf{Fragmented prompt-handling and problem-solving in LLM output\text{*}:} LLM responses tend to be fragmented as the models handle prompts in isolated pieces rather than forming a holistic understanding, thus lacking bidirectional and closed-looped interactions and contextual understanding.
    
    \item \textbf{Lagging governance and public education:} LLMs create new challenges to AI literacy and technology regulation, while governance and public education on this emerging technology are still lagging.
\end{itemize}


\section{Discussion}

Our findings highlight the additional challenges LLMs introduce into the information ecosystem. Low-quality information is not limited to factual inaccuracy; it can manifest through nuanced forms of harm, such as omission or misprioritization, that fact-checking alone cannot address. These subtler failures can still influence how people interpret, prioritize, and act on information, particularly in high-stakes contexts like health.

At the same time, LLM affordances are fundamentally different from those of traditional information sources. Their fluency, adaptability, and interactive nature can shift how people search for, evaluate, and depend on information, which potentially can reshape habits, expectations, and trust in ways we do not yet fully understand. These dynamics demand that the community expand both its conceptual boundaries and methodological approaches, moving beyond a singular focus on truthfulness to encompass the full spectrum of quality, context, and user impact.

By documenting types of LLM-generated low-quality information and identifying LLM distinctive characteristics, we call for more attention and collaborative efforts to develop frameworks, interventions, and governance strategies that can address these emerging risks. We believe the typologies offered by this work can serve as beginning steps and basic introductory material for the general public to understand the mechanisms and constraints of the technology, as well as for interdisciplinary stakeholders with varying levels of AI knowledge to support informed discussions about what informational harms may emerge and how they might be mitigated.


\newpage

\bibliographystyle{ACM-Reference-Format}
\bibliography{references}

\appendix

\end{document}
\endinput